\documentclass[
    ,final            
  ]
  {aipproc}

\layoutstyle{6x9}

\begin{document}

\title{Hadroproduction of $D$ and $B$ mesons in a massive VFNS}



\author{B.\ A.\ Kniehl, G.\ Kramer, I.\ Schienbein}{
  address={II.\ Institut f\"ur Theoretische Physik, Universit\"at Hamburg,
Luruper Chaussee 149,\\ 22761 Hamburg, Germany}
}

\author{H.\ Spiesberger}{
  address={Institut f\"ur Physik, Johannes-Gutenberg-Universit\"at,
Staudinger Weg 7, 55099 Mainz, Germany}
}
%

\begin{abstract}
We present
a calculation of the next-to-leading order cross section for the
inclusive hadroproduction of $D$ and $B$ mesons as a function
of the transverse momentum and the rapidity
in a massive variable flavor number scheme.
We compare our numerical results with recent data 
from the CDF Collaboration at the Fermilab Tevatron 
for the production of $D^0$, $D^{\star +}$, $D^+$, and $D_s^+$ mesons
at center-of-mass energy
$\sqrt{S} = 1.96\ {\rm TeV}$ and find reasonably good agreement with 
the measured cross sections.
\end{abstract}

\maketitle



Various approaches for next-to-leading order (NLO) calculations
in perturbative QCD 
have been applied to one-particle inclusive
hadroproduction of $D$ or $B$ mesons.
For definiteness, we shall consider here $D$ mesons.
However, all results can easily be carried over to any other
heavy-flavored hadron.

A basic approach is the fixed flavor number 
scheme (FFNS) 
\cite{FO},
in which the number of active flavors in the initial state is fixed 
to $n_f = 3$ and the charm quark appears only in the final state. 
The charm mass $m$ is
explicitly taken into account together with the transverse momentum 
$p_T$ of the observed meson. 
In this scheme the charm mass acts as a cutoff for the 
initial- and final-state collinear singularities and
collinear logarithms $\ln(p_T^2/m^2)$ are kept in the
hard scattering cross sections.
However, for $p_T \gg m$, these logarithms become large 
and spoil the convergence of the perturbation series.

Therefore, in the regime $p_T \gg m$, it is more appropriate
to treat charm quarks like massless partons and to absorb
the collinear logarithms into scale dependent parton distribution
functions (PDFs) and fragmentation functions (FFs).
As is well-known, by this procedure 
the large logarithms $\ln(p_T^2/m^2)$ are summed via the 
DGLAP evolution equations and
the hard scattering cross sections are finite (infrared safe)
in the limit $m \rightarrow 0$. 
If the power-like charm mass terms ${\cal O}(m^2/p_T^2)$ are neglected 
this is just the conventional
parton model or zero-mass variable flavor number scheme (ZM-VFNS).
Usually, in the ZM-VFNS the charm mass is neglected from the beginning
and the collinear singularities appear in dimensional regularization
as poles in $\epsilon$ where $d=4- 2\epsilon$ is the number of 
space-time dimensions. Conventionally, these poles are removed in
the modified-minimal-subtraction ($\overline{MS}$) scheme.
If, on the other hand, the collinear singularities have been 
regularized with help of a mass $m$ it is necessary also to 
subtract finite terms along with the collinear logarithms $\ln m^2$ in
order to recover the hard scattering cross sections in the
$\overline{MS}$ scheme.

On top of these two basic approaches,
schemes have been devised which combine the two features, non-zero
charm mass and resummation of $\ln(p_T^2/m^2)$-terms.
One such scheme, which has been applied already to 
inclusive charmed meson production 
for the Tevatron experiment is the so-called fixed-order
next-to-leading-logarithm (FONLL) scheme. 
This scheme smoothly interpolates between the
traditional cross section in the FFNS
and a suitably modified cross section in the
ZM-VFNS approach with perturbative FFs 
with the help of a $p_T$ dependent weight function 
\cite{Cacciari:1998it,Cacciari:2003zu}. 
In both
non-zero-charm-mass approaches, FFNS and FONLL, the theoretically
calculated cross sections are convoluted with a 
scale-independent non-perturbative FF 
extracted from $e^+ e^-$ data describing the transition from the
produced charm quark to the observed $D$ meson.

Recently, a general mass variable flavor number scheme (GM-VFNS)
has been worked out by us 
\cite{Kniehl:2004fy,Kniehl:2005mk,Schienbein:2004ah,Schienbein:2003et}
which is closely related to the ZM-VFNS, but keeps all $m^2/p_T^2$ terms 
in the hard-scattering cross sections
in order to achieve better accuracy in the intermediate 
region $p_T \geq m$. 
The massive hard scattering cross sections have been constructed
in a way that the conventional hard scattering cross sections
in the $\overline{MS}$ scheme are recovered in the limit $p_T \to \infty$
(or $m \to 0$).
The requirement to adjust the massive theory to the 
ZM-VFNS with $\overline{MS}$ subtraction is necessary since all commonly 
used PDFs and FFs for heavy flavors are defined in this particular scheme. 
In this sense this subtraction scheme is a consistent extension of the 
conventional ZM-VFNS for including charm-quark mass effects. 
It should be noted that our implementation of a GM-VFNS is similar 
to the ACOT scheme
which has been extended to 1-particle inclusive production of
$B$ mesons a few years ago \cite{Olness:1997yc}.
There are small differences concerning the 
collinear subtraction terms \cite{Kniehl:2005mk}. 
Further, in \cite{Olness:1997yc}, the resummation of
the final state collinear logarithms has been performed only to
leading logarithmic accuracy.


\begin{figure*}[t]
\begin{tabular}{llll}
{\parbox{3.5cm}{
\hspace*{-1.0cm}
\includegraphics[height=.25\textheight]{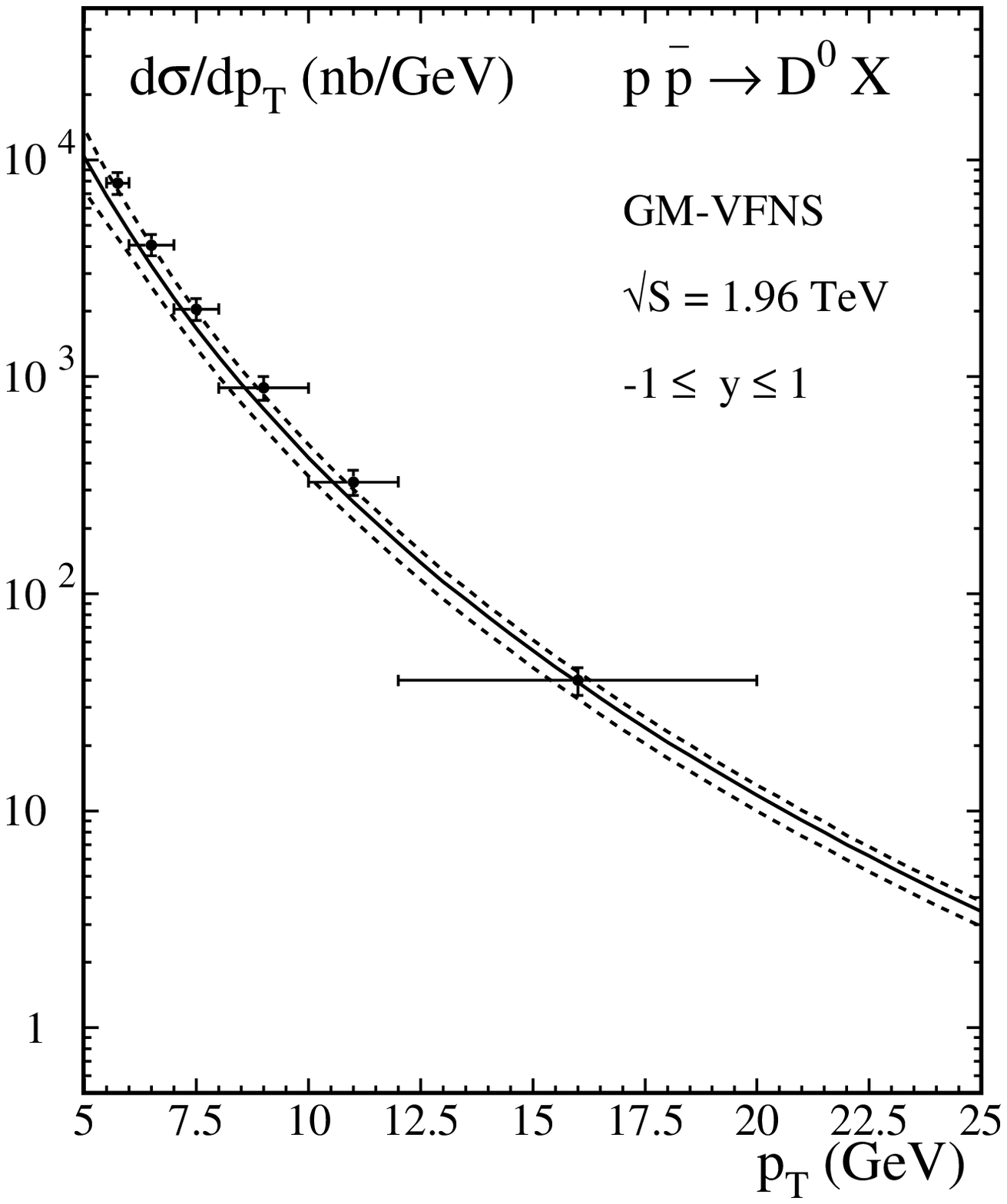}
}} &
{\parbox{3.5cm}{
\hspace*{-0.8cm}
\includegraphics[height=.25\textheight]{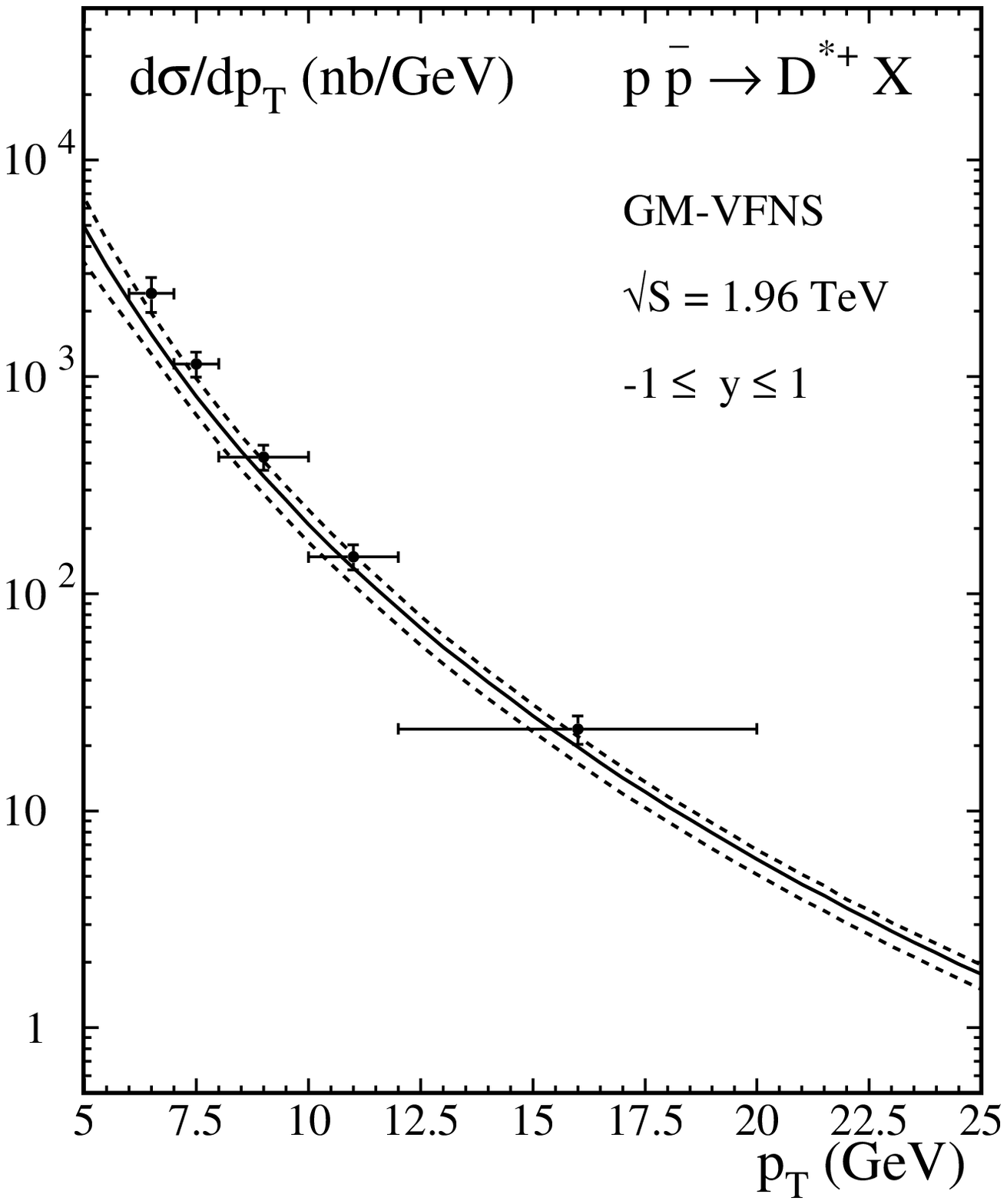}
}} &
{\parbox{3.5cm}{
\hspace*{-0.6cm}
\includegraphics[height=.25\textheight]{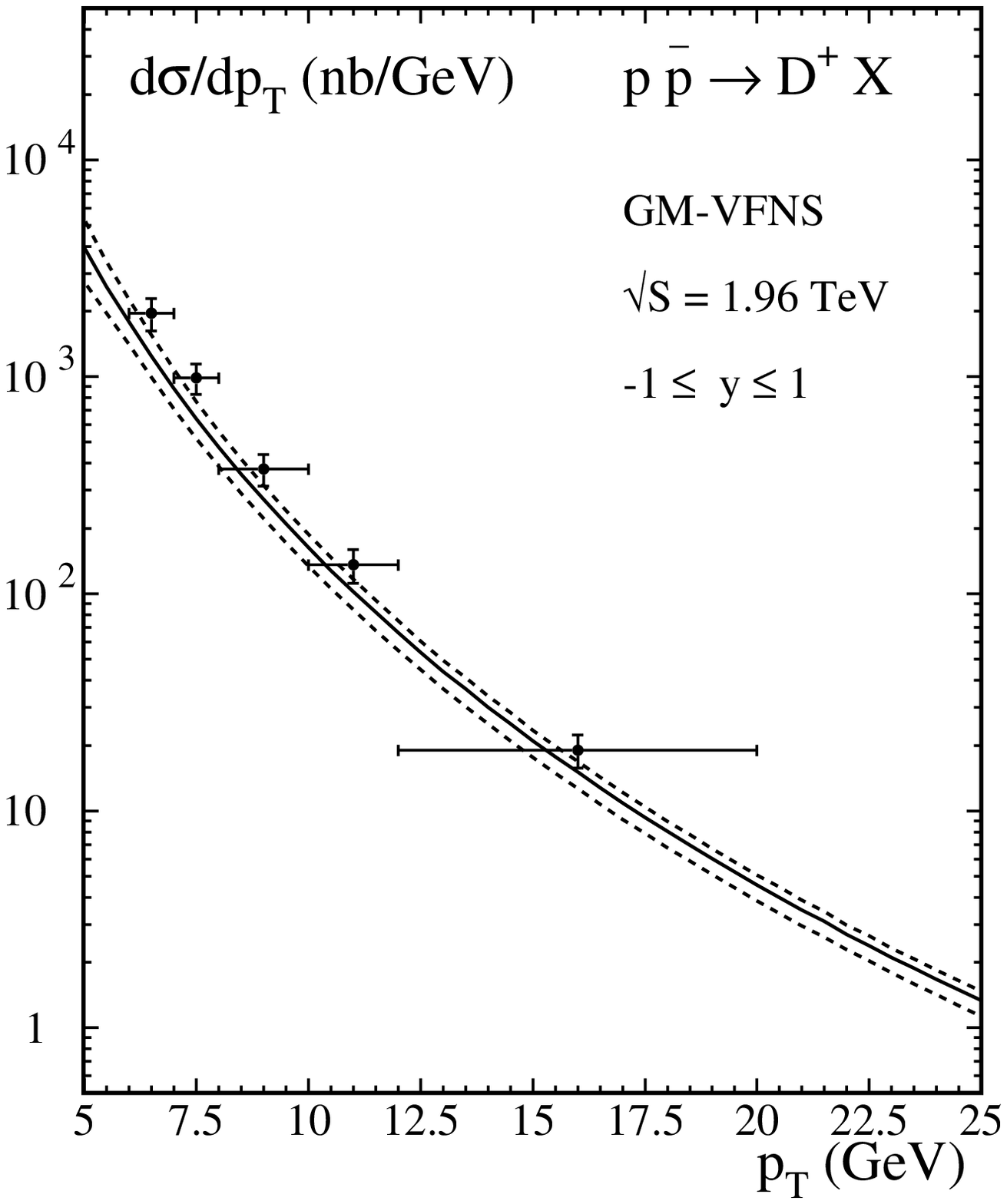}
}} &
{\parbox{3.5cm}{
\hspace*{-0.4cm}
\includegraphics[height=.25\textheight]{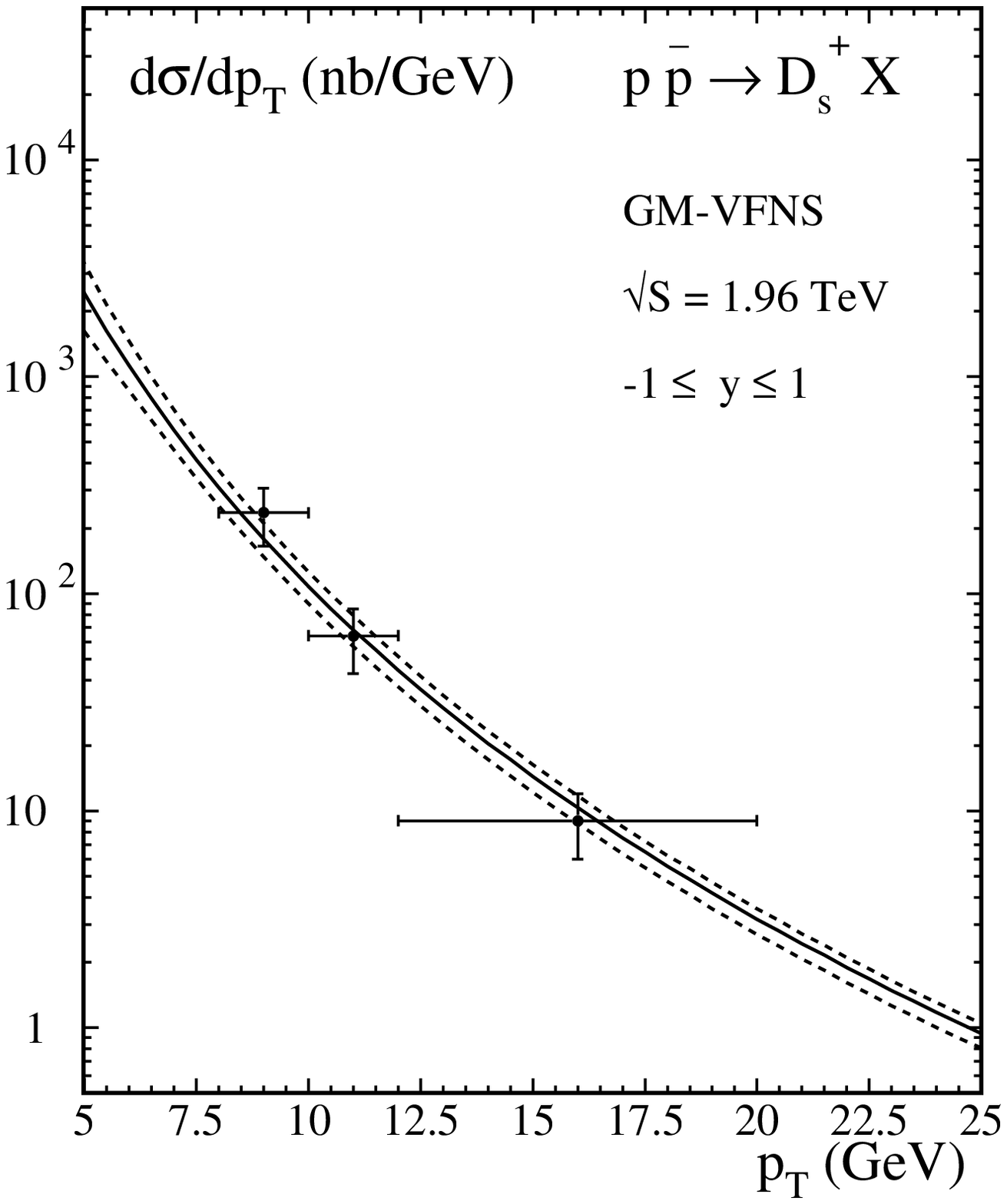}
}}
\end{tabular}
 \caption{
QCD predictions for one-particle inclusive production of
charmed mesons, $X_c = D^0, D^{\star +}, D^+, D_s^+$, at 
the Tevatron Run II.
In each case,
the results are shown for the average 
of the observed meson with its antiparticle
$(X_c + \overline{X}_c)/2$.
The solid lines have been obtained with 
$\mu_{\rm R}=\mu_{\rm F}=\mu_{\rm F}^\prime = m_T$.
The upper and lower dashed curves represent the maximum and
minimum cross sections found by varying $\mu_{\rm R}$, $\mu_{\rm F}$,
and $\mu_{\rm F}^\prime$ independently within a factor of 2 up and
down relative to the central values
while keeping their ratios
$0.5 \le \mu_{\rm F}/\mu_{\rm R}, \mu_{\rm F}^\prime/\mu_{\rm R}, 
\mu_{\rm F}/\mu_{\rm F}^\prime \le 2$.
CDF data \protect\cite{Acosta:2003ax} are shown for comparison.
}
\label{fig:fig1}
\end{figure*}


\begin{figure*}[t]
\begin{tabular}{llll}
{\parbox{3.5cm}{
\hspace*{-1.0cm}
\includegraphics[height=.25\textheight]{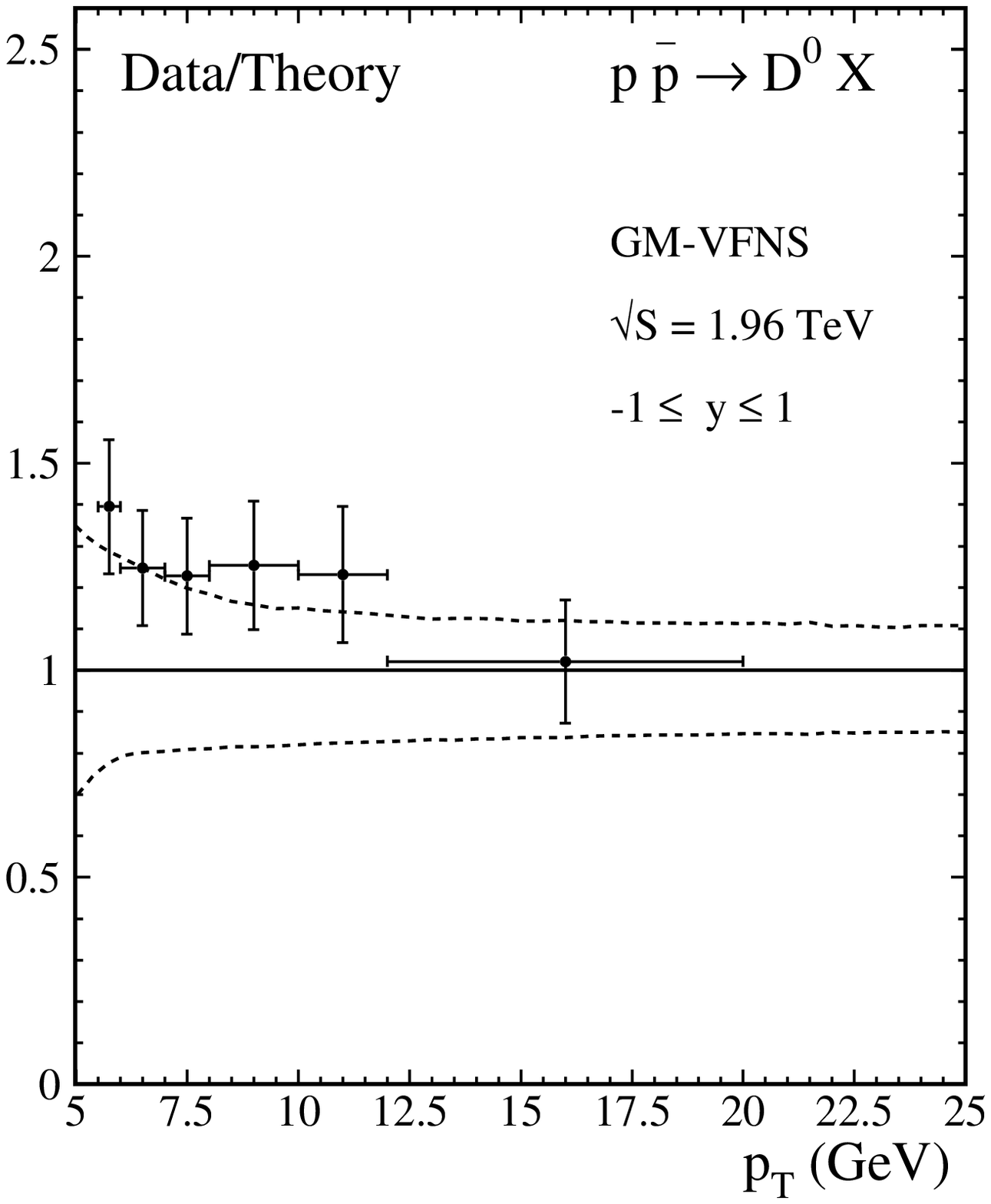}
}} &
{\parbox{3.5cm}{
\hspace*{-0.8cm}
\includegraphics[height=.25\textheight]{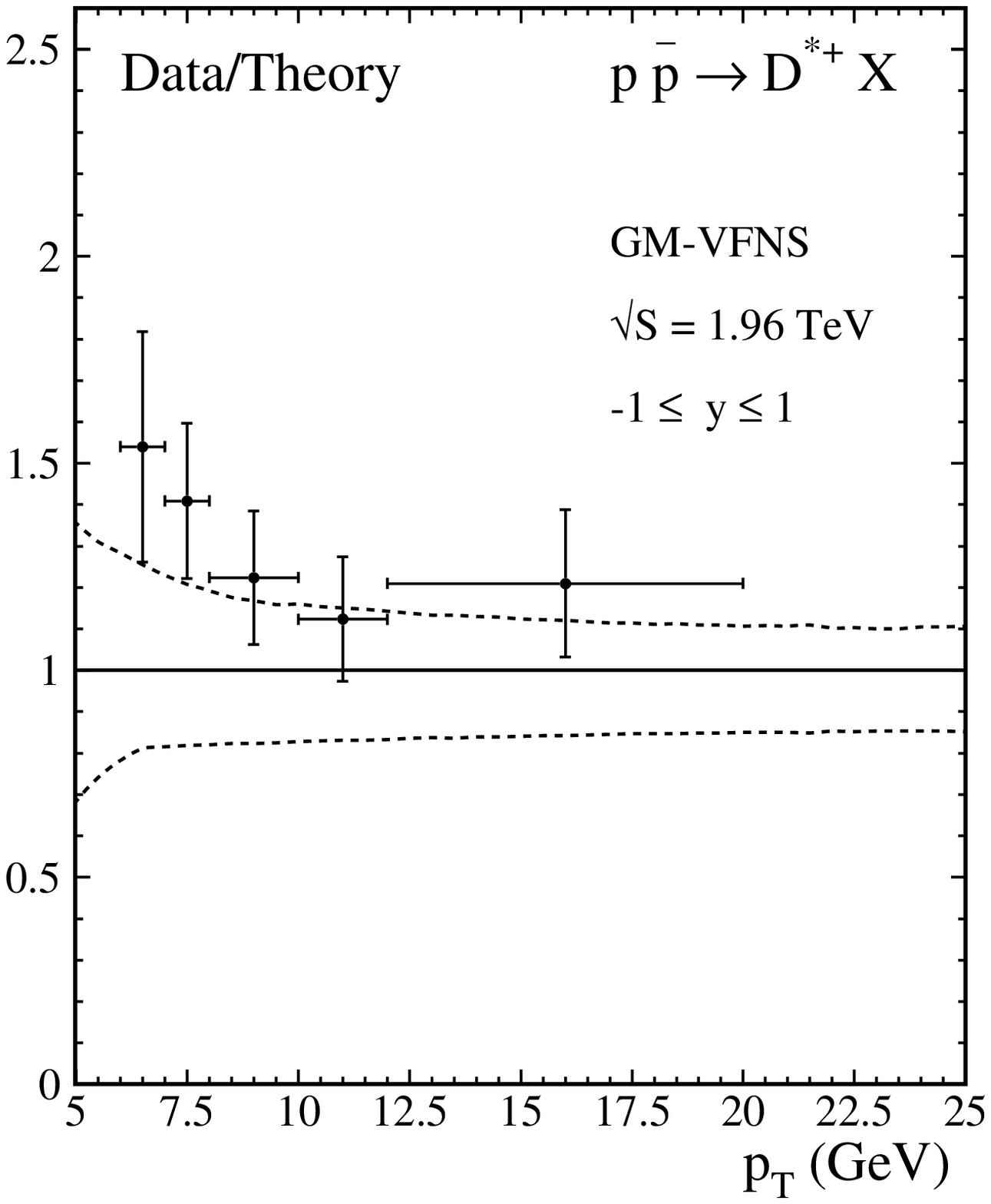}
}} &
{\parbox{3.5cm}{
\hspace*{-0.6cm}
\includegraphics[height=.25\textheight]{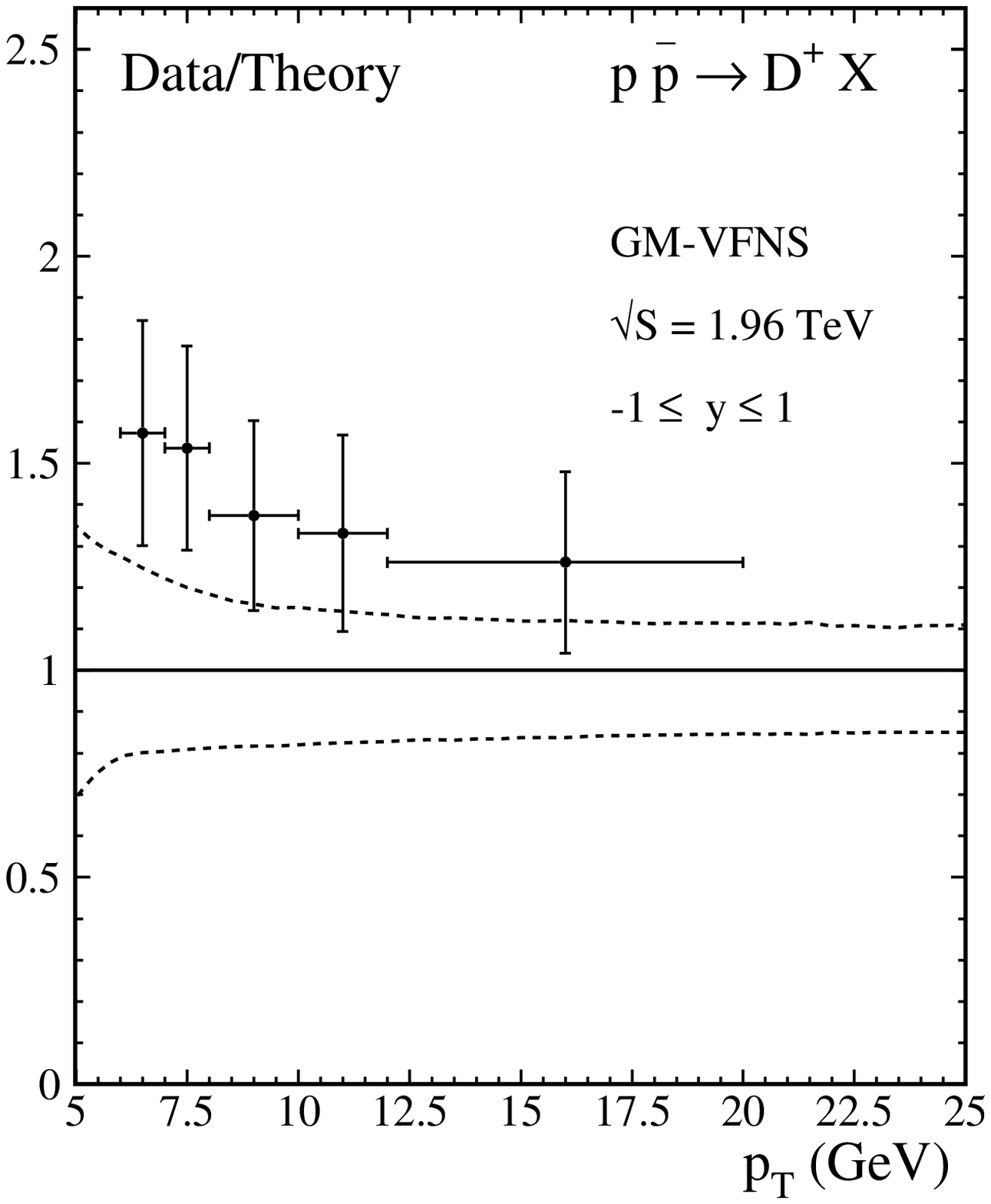}
}} &
{\parbox{3.5cm}{
\hspace*{-0.4cm}
\includegraphics[height=.25\textheight]{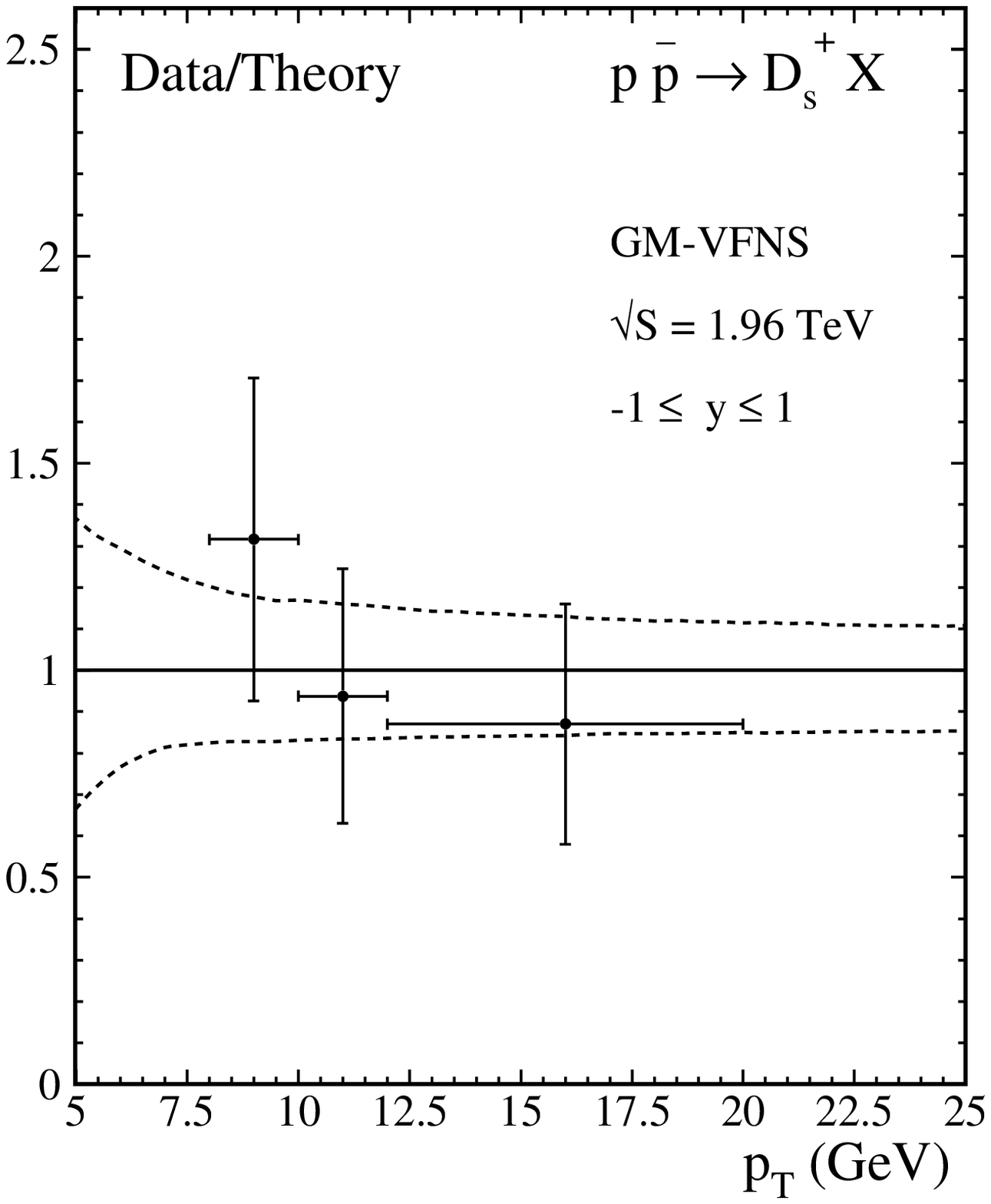}
}}
\end{tabular}
 \caption{
Ratios of experimental results and our central theoretical predictions
(solid lines in Fig.\ \protect\ref{fig:fig1}).
In addition, the theoretical uncertainty bands are shown, obtained
as the ratio of the upper (lower) QCD prediction and the central
curve.
}
\label{fig:fig2}
\end{figure*}


To calculate the cross section $d^2\sigma/dp_Tdy$ for the reactions 
$p+\bar{p} \rightarrow D + X$, 
FFs are needed which describe the fragmentation of 
the charm quarks, the light quarks, and the gluon into the 
observed $D$ mesons. 
Fragmentation functions for the $D^{\star}$ meson have been 
extracted at leading and next-to-leading order already some time 
ago \cite{Binnewies:1998xq}, using experimental data from the 
OPAL \cite{Ackerstaff:1997ki} and ALEPH \cite{Barate:1999bg} 
collaborations at LEP1. 
Recently, using the same procedure as in \cite{Binnewies:1998xq},
also the FFs for $X_c = D^{0}, D^{+}, D_s^{+}, \Lambda_c$
have been determined \cite{Kniehl:2005de} using OPAL data for
$e^+e^- \rightarrow X_c + X$ \cite{Alexander:1996wy}.
In Refs.\ \cite{Binnewies:1998xq,Kniehl:2005de} 
the fits for the $X_c$ FFs 
have been performed using
a starting scale $\mu_0=2 m$
for the gluon and the $u, d, s$ and $c$ quarks and their antiquarks, 
while $\mu_0=2 m_b$ ($m_b = 5\ {\rm GeV}$) was
chosen for the FFs of the bottom quark and antiquark.
The FFs of the gluon and the first three flavors were assumed 
to be zero at this $\mu_0$.
At larger scale $\mu$, these FFs are generated through the usual 
DGLAP evolution.
Since the effect of the gluon FF is important
at Tevatron energies
as was found for $D^{\star}$ production in \cite{Kniehl:2004fy} 
we decided to repeat the fits for the $X_c$ FFs with the lower 
starting scales $\mu_0=m$ and $\mu_0=m_b$, respectively.
This changes the FFs of the $c$ quark only marginally but has a sizable 
effect on the gluon FF. The details of these new FFs will be presented 
elsewhere \cite{ffs}.

Next we show our predictions for the cross sections $d\sigma/dp_T$ for 
$D^0, D^{\star+}, D^{+}$ and $D_s^{+}$ production obtained 
in the GM-VFNS. 
For a comparison with the ZM-VFNS we refer
to Ref.\ \cite{heralhc}.
The partonic cross sections are convoluted with the 
(anti-)proton PDFs and the FFs for $c \rightarrow X_c$, 
$u,d,s \rightarrow X_c$ and $g \rightarrow X_c$. 
We use CTEQ6M PDFs \cite{Pumplin:2002vw} and the FF sets for 
$D^0$, $D^{\star+}$, $D^{+}$ and $D_s^{+}$ from \cite{ffs}.
\footnote{It should be noted that the results presented at the
DIS05 have been obtained with the FFs from 
\protect\cite{Binnewies:1998xq,Kniehl:2005de}.}
Results are shown for the average of the observed $X_c$ mesons with
their antiparticles.
We consider $d\sigma/dp_T$ at
$\sqrt{S} = 1.96\ {\rm TeV}$ as a function 
of $p_T$  with $y$ integrated over the range $ -1.0 <  y  < 1.0$. 
For the charm mass we take $m = 1.5\ {\rm GeV}$ and 
evaluate $\alpha_s^{(n_f)}(\mu_R)$ with $n_f=4$ and scale parameter 
$\Lambda^{(4)}_{\overline{MS}} = 328\ {\rm MeV}$, corresponding to 
$\alpha_s^{(5)}(m_Z) = 0.1181$. 
The results are presented in Figs.\ \ref{fig:fig1} and \ref{fig:fig2}. 
The solid lines correspond to the central scale choice 
$\mu_R=\mu_F=\mu_F'=m_T=(p_T^2+m^2)^{1/2}$,
where $\mu_R$ is the renormalization, $\mu_F$ the initial-state 
and $\mu_F'$ the 
final-state factorization scale, respectively. 
To investigate the scale variation of our predictions, 
we independently 
vary the renormalization and factorization scales by a factor of two:
$0.5 \le \mu_{\rm R}/m_T, \mu_{\rm F}/m_T, \mu_{\rm F}^\prime/m_T \le 2$
while keeping their ratios
$0.5 \le \mu_F/\mu_R, \mu_F'/\mu_R, \mu_F/\mu_F' \le 2$ \cite{heralhc}.
Our theoretical results are compared with the 
experimental data from CDF \cite{Acosta:2003ax}. 
As can be seen, the data are in good agreement with 
the upper curve of the uncertainty band whereas
they are a factor of about $1.5$($1.2$) 
above our central prediction at low(high) $p_T$. 

Residual sources of theoretical uncertainty include the variations of the
charm mass and the assumed PDF and FF sets. 
A variation of the value of the charm-quark mass does not contribute 
much to the theoretical uncertainty.
Also the use of other up-to-date NLO proton PDF sets produces only minor
differences. 
Concerning the choice of the NLO FF sets we obtain 
results reduced by a factor of $1.2$--$1.3$
when we use the NLO sets obtained
by fitting with the initial scale choice $\mu_0=2 m, 2 m_b$. 

In conclusion, we have presented a NLO perturbative QCD calculation 
of $D$ meson production at the Tevatron in a GM-VFNS 
\cite{Kniehl:2004fy,Kniehl:2005mk}
which provides the best description of these experimental results 
obtained so far.
It completes earlier work in this scheme on $D$ meson production
in $\gamma \gamma$ and $\gamma p$ collisions 
\cite{KS}.
This approach will be applied next to $B$ meson production at the
Tevatron. Furthermore, it is planned to extend this scheme
to heavy meson production in deep inelastic scattering.

\begin{theacknowledgments}
This work was supported in part by the Bundesministerium f\"ur Bildung
und Forschung through Grant No.\ 05 HT4GUA/4.
The work of I.\ S.\ was supported by DESY.
\end{theacknowledgments}





\begin{thebibliography}{22}
\providecommand{\eprint}[2][]{\url{#2}}

\bibitem{FO}
P.~Nason, S.~Dawson, and R.~K. Ellis, \emph{Nucl. Phys.}, 
\textbf{B303}, 607 (1988); \textbf{B327}, 49 (1989); 
\textbf{B335}, 260(E) (1990);
W.~Beenakker, H.~Kuijf, W.~L. van Neerven, and J.~Smith, 
\emph{Phys. Rev.}, \textbf{D40}, 54 (1989);
W.~Beenakker, W.~L. van Neerven, R.~Meng, G.~A. Schuler, and J.~Smith, 
\emph{Nucl. Phys.}, \textbf{B351}, 507 (1991);
I.~Bojak, and M.~Stratmann, \emph{Phys. Rev.}, \textbf{D67}, 
034010 (2003).

\bibitem{Cacciari:1998it}
M.~Cacciari, M.~Greco, and P.~Nason, \emph{JHEP}, \textbf{05}, 007 (1998).

\bibitem{Cacciari:2003zu}
M.~Cacciari, and P.~Nason, \emph{JHEP}, \textbf{09}, 006 (2003).

\bibitem{Kniehl:2004fy}
B.~A. Kniehl, G.~Kramer, I.~Schienbein, and H.~Spiesberger, 
\emph{Phys. Rev.}, \textbf{D71}, 014018 (2005).

\bibitem{Kniehl:2005mk}
B.~A. Kniehl, G.~Kramer, I.~Schienbein, and H.~Spiesberger, 
\emph{Eur. Phys. J.}, \textbf{C41}, 199 (2005).


\bibitem{Schienbein:2004ah}
I.~Schienbein, \eprint{hep-ph/0408036}.

\bibitem{Schienbein:2003et}
I.~Schienbein, {Open heavy-flavour photoproduction at NLO}, 
{Proceedings of the Ringberg Workshop, 
{\it New Trends in HERA Physics 2003}, edited by
  G.\ Grindhammer, B.\ A.\ Kniehl, G.\ Kramer and W.\ Ochs, World Scientific,
  2004, p.\ 197}.

\bibitem{Olness:1997yc}
F.~I. Olness, R.~J. Scalise, and W.-K. Tung, \emph{Phys. Rev.}, 
\textbf{D59}, 014506 (1999).

\bibitem{Acosta:2003ax}
D.~Acosta, et~al., \emph{Phys. Rev. Lett.}, \textbf{91}, 241804 (2003).

\bibitem{Binnewies:1998xq}
J.~Binnewies, B.~A. Kniehl, and G.~Kramer, \emph{Phys. Rev.}, 
\textbf{D58}, 014014 (1998).

\bibitem{Ackerstaff:1997ki}
K.~Ackerstaff, et~al., \emph{Eur. Phys. J.}, \textbf{C1}, 
439 (1998).

\bibitem{Barate:1999bg}
R.~Barate, et~al., \emph{Eur. Phys. J.}, \textbf{C16}, 
597 (2000).

\bibitem{Kniehl:2005de}
B.~A. Kniehl, and G.~Kramer, \emph{Phys. Rev.}, \textbf{D71}, 
094013 (2005).

\bibitem{Alexander:1996wy}
G.~Alexander, et~al., \emph{Z. Phys.}, \textbf{C72}, 1 (1996).

\bibitem{ffs}
B.~A. Kniehl, G.~Kramer, I.~Schienbein, and H.~Spiesberger, {in preparation}.

\bibitem{heralhc}
{See the Proceedings of the HERA-LHC Workshop}.

\bibitem{Pumplin:2002vw}
J.~Pumplin, et~al., \emph{JHEP}, \textbf{07}, 012 (2002).

\bibitem{KS}
G.~Kramer, and H.~Spiesberger, \emph{Eur. Phys. J.}, \textbf{C22}, 
289 (2001); \textbf{C28}, 495 (2003); \textbf{C38}, 309 (2004). 

\end{thebibliography}

\end{document}